\documentclass{article}
\usepackage[preprint]{colm2026_conference}

\usepackage{microtype}
\usepackage{hyperref}
\usepackage{url}
\usepackage{booktabs}
\usepackage{tabularx}
\usepackage{array}
\usepackage{amsmath}
\usepackage{amssymb}
\usepackage{tikz}
\usepackage{xcolor}
\usepackage{xspace}

\usetikzlibrary{arrows.meta,calc,fit,positioning}

\definecolor{darkblue}{rgb}{0,0,0.5}
\definecolor{cvblue}{HTML}{2F6B9A}
\definecolor{cvgreen}{HTML}{3F8060}
\definecolor{cvamber}{HTML}{A56A16}
\definecolor{cvred}{HTML}{9B4A53}
\definecolor{lightblue}{HTML}{EAF3F8}
\definecolor{lightgreen}{HTML}{ECF5EF}
\definecolor{lightamber}{HTML}{FAF1E3}
\definecolor{lightred}{HTML}{F8ECEE}
\definecolor{lightgray}{HTML}{F3F4F5}
\hypersetup{
  colorlinks=true,
  citecolor=darkblue,
  linkcolor=darkblue,
  urlcolor=darkblue,
  pdftitle={The Evolving Bottleneck in Speech Generation: Interface Co-design and Staged Alignment from CosyVoice to Qwen-Audio-3.0-TTS},
  pdfauthor={Qian Chen, Xiangang Li, Xiang Lv, Han Zhao, Tianyu Zhao}
}

\newcolumntype{Y}{>{\raggedright\arraybackslash}X}
\newcolumntype{P}[1]{>{\raggedright\arraybackslash}p{#1}}
\newcommand{\system}{Qwen-Audio-3.0-TTS\xspace}
\newcommand{\cvone}{CosyVoice\xspace}
\newcommand{\cvtwo}{CosyVoice~2\xspace}
\newcommand{\cvthree}{CosyVoice~3\xspace}
\newcommand{\LM}{\textsc{lm}\xspace}
\newcommand{\FM}{\textsc{fm}\xspace}

\title{The Evolving Bottleneck in Speech Generation:\\
Interface Co-design and Staged Alignment from CosyVoice to Qwen-Audio-3.0-TTS}

\author{Qian Chen \quad Xiangang Li \quad Xiang Lv \quad Han Zhao \quad Tianyu Zhao\\
Alibaba Token Foundry\\
\textit{Equal contribution; alphabetical by last name.}}

\begin{document}
\maketitle

\begin{abstract}
Speech synthesis systems are commonly narrated as a sequence of larger models, better
tokenizers, and broader data. This technical retrospective offers a different account of
the CosyVoice lineage, from CosyVoice through CosyVoice~2 and CosyVoice~3 to
Qwen-Audio-3.0-TTS: progress came from repeatedly relocating the system's dominant
bottleneck. Across the lineage, a stable decomposition separates an autoregressive
language model that plans speech from a flow-matching model that renders acoustics.
What changes is the contract between them. CosyVoice establishes supervised semantic
tokens as a content-aligned interface; CosyVoice~2 makes that interface causally
available for streaming and removes the utterance-level speaker embedding from the
language model;
CosyVoice~3 improves the learnability and coverage of the interface through
multitask supervision, scaling, and differentiable reward optimization; and
Qwen-Audio-3.0-TTS reduces token rate, conditions its renderer on continuous language-model
hidden states instead of token embeddings, and progressively aligns the coupled system.
We formalize this history through four
interface dimensions---representation, ownership, availability, and gradient reach---and
separate within-paper evidence from cross-paper comparison. The resulting synthesis
connects discrete autoregressive, continuous non-autoregressive, hybrid, and continuous
autoregressive speech-generation paradigms, and yields practical principles for
diagnosing and training modular speech generators.
\end{abstract}

\section{Introduction}

Modern speech generation must optimize a crowded frontier: linguistic correctness,
speaker fidelity, prosodic naturalness, audio quality, multilingual and dialectal
coverage, controllability, latency, long-form stability, and robustness to imperfect
reference speech. These objectives do not fail in the same module, and improving one can
expose a new constraint elsewhere. A representation rich enough for acoustic
reconstruction may be difficult for an autoregressive model to predict. A semantic
tokenizer may enable stable planning but discard cues required for expressive
realization. A strong offline model may become unsuitable for streaming because its
acoustic path requires future context. A large training mixture may broaden coverage
while leaving a gap between token accuracy and perceptual quality.

This paper studies those shifts through the lineage formed by
\cvone~\citep{du2024cosyvoice}, \cvtwo~\citep{du2024cosyvoice2},
\cvthree~\citep{du2025cosyvoice3}, and \system~\citep{xiang2026qwenaudio}.
The releases share enough structure to support a structured conceptual comparison:
each uses an autoregressive language model (\LM) to produce a compact speech plan and a
flow-matching model (\FM) to reconstruct continuous acoustic features. The newest
paper explicitly describes its system as building on CosyVoice~2 and CosyVoice~3,
despite the change in product name. This continuity lets us ask a more useful question
than ``which release is best?'': \emph{what was targeted after the previous constraint
had been addressed?}

Our answer is the \emph{evolving-bottleneck thesis}. The dominant constraint moves
through four stages:
\begin{enumerate}
  \item \textbf{representation}: learning speech units that align with content and are
  predictable from text;
  \item \textbf{availability and ownership}: making those units causal enough for
  streaming and deciding which module should receive explicit speaker conditioning;
  \item \textbf{learnability and coverage}: scaling data and model capacity while
  keeping a compact interface trainable and rewardable;
  \item \textbf{interface capacity and coordination}: allowing the renderer to access
  information beyond token IDs and letting acoustic losses shape upstream
  representations.
\end{enumerate}

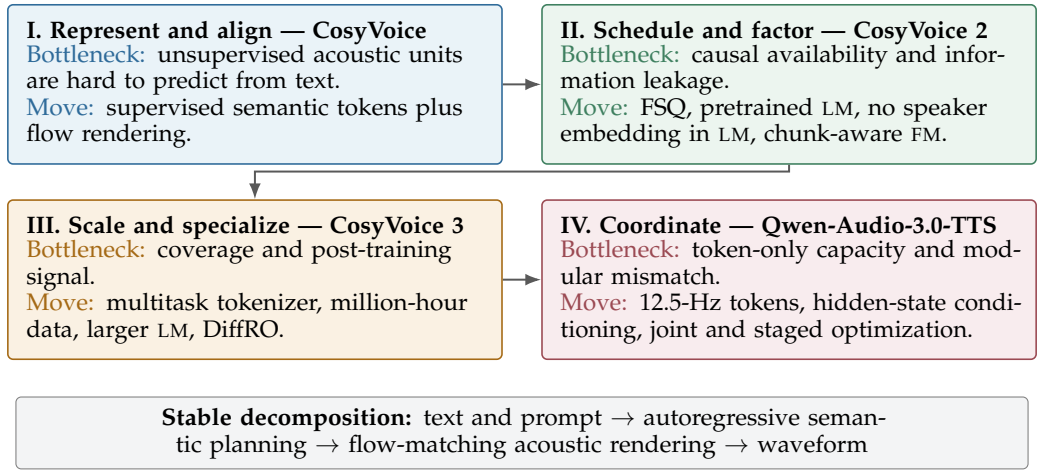
\begin{figure}[t]
\centering
\begin{tikzpicture}[
  font=\small,
  box/.style={rounded corners=2pt, draw, line width=0.6pt, align=left,
    inner xsep=7pt, inner ysep=6pt, text width=2.38in, minimum height=0.83in},
  arr/.style={-{Latex[length=2.1mm]}, line width=0.75pt, draw=black!65}
]
  \node[box, draw=cvblue, fill=lightblue] (a) at (0,0) {
    \textbf{I. Represent and align --- CosyVoice}\\[-1pt]
    \textcolor{cvblue}{Bottleneck:} unsupervised acoustic units are hard to predict
    from text.\\
    \textcolor{cvblue}{Move:} supervised semantic tokens plus flow rendering.
  };
  \node[box, draw=cvgreen, fill=lightgreen, right=0.20in of a] (b) {
    \textbf{II. Schedule and factor --- CosyVoice 2}\\[-1pt]
    \textcolor{cvgreen}{Bottleneck:} causal availability and information leakage.\\
    \textcolor{cvgreen}{Move:} FSQ, pretrained \LM, no speaker embedding in \LM,
    chunk-aware \FM.
  };
  \node[box, draw=cvamber, fill=lightamber, below=0.18in of a] (c) {
    \textbf{III. Scale and specialize --- CosyVoice 3}\\[-1pt]
    \textcolor{cvamber}{Bottleneck:} coverage and post-training signal.\\
    \textcolor{cvamber}{Move:} multitask tokenizer, million-hour data, larger \LM,
    DiffRO.
  };
  \node[box, draw=cvred, fill=lightred, right=0.20in of c] (d) {
    \textbf{IV. Coordinate --- Qwen-Audio-3.0-TTS}\\[-1pt]
    \textcolor{cvred}{Bottleneck:} token-only capacity and modular mismatch.\\
    \textcolor{cvred}{Move:} 12.5-Hz tokens, hidden-state conditioning, joint and
    staged optimization.
  };
  \draw[arr] (a.east) -- (b.west);
  \draw[arr] (b.south) -- ++(0,-0.09) -| (c.north);
  \draw[arr] (c.east) -- (d.west);
  \coordinate (gridbottom) at ($(c.south)!0.5!(d.south)$);
  \node[draw=black!50, fill=lightgray, rounded corners=2pt, inner xsep=8pt,
    inner ysep=4pt, align=center, text width=5.05in, below=0.19in of gridbottom]
    (stable) {
    \textbf{Stable decomposition:}
    text and prompt $\rightarrow$ autoregressive semantic planning
    $\rightarrow$ flow-matching acoustic rendering $\rightarrow$ waveform
  };
\end{tikzpicture}
\caption{The lineage as a sequence of bottleneck relocations. The planner--renderer
decomposition remains stable; the interface and its training contract evolve. Each
release targets a different constraint; the arrows encode our retrospective
interpretation rather than a controlled cross-release experiment.}
\label{fig:lineage}
\end{figure}

Figure~\ref{fig:lineage} summarizes this progression. The key point is not that later
systems discard discrete tokens. In \system, tokens remain essential for semantic
planning, efficient autoregressive decoding, and direct token-domain optimization. What
ends is their role as the renderer's direct conditioning interface: the semantic-token
prediction head is retained, while continuous hidden states condition the \FM.
Progressive training coordinates the modules without sacrificing their useful
factorization.

This retrospective makes three contributions. First, it introduces an interface
contract with four dimensions---representation, ownership, availability, and gradient
reach---for comparing modular speech generators. Second, it reconstructs the
CosyVoice--Qwen-Audio lineage as a sequence of bottleneck relocations, grounding each
mechanistic claim in within-paper ablations or explicit architecture changes. Third, it
places the lineage within the broader design space of discrete autoregressive,
continuous non-autoregressive, hybrid, and continuous autoregressive synthesis, and
derives an evaluation protocol that avoids misleading cross-paper leaderboards.

\section{Scope, evidence, and analytical framework}
\label{sec:framework}

\paragraph{A technical retrospective, not a synthetic leaderboard.}
The four source papers were produced at different times, with changing data, checkpoints,
evaluation sets, recognizers, speaker encoders, and deployment targets. We therefore do
not interpret cross-paper score differences as controlled causal effects. Architecture
claims are supported by the documented model designs; mechanistic claims are anchored in
ablations performed within a source paper. Cross-release numbers are reported only when
the newest work explicitly evaluates earlier and newer tokenizers in the same table.
This evidence discipline matters: the bottleneck sequence is our mechanistic
interpretation of the published evidence, not an assertion that publication order alone
proves causality.

\paragraph{The interface contract.}
Let $x$ denote text and prompt context, $z_{1:T}$ a discrete speech plan, $h_{1:T}$ the
\LM hidden sequence, and $y$ continuous acoustic features. A cascade implements
\begin{equation}
 p_\theta(z\mid x)\quad\text{and}\quad p_\phi(y\mid e(z),x),
 \label{eq:cascade}
\end{equation}
where $e(z)$ denotes token embeddings. In the coupled system, token prediction remains
an auxiliary semantic objective, but renderer conditioning changes to
\begin{equation}
 p_\phi(y\mid h,x), \qquad
 \mathcal{L}=\mathcal{L}_{\mathrm{token}}+
 \lambda\mathcal{L}_{\mathrm{flow}},
 \label{eq:dual}
\end{equation}
with $\nabla_\theta\mathcal{L}_{\mathrm{flow}}\neq 0$ during joint training.
We characterize the planner--renderer contract as
\begin{equation}
 \mathcal{I}=(R,O,A,G),
 \end{equation}
where $R$ is the representation passed across the boundary, $O$ assigns factors such as
content and speaker identity to modules, $A$ specifies when interface values become
available at inference, and $G$ specifies which objectives can send gradients across the
boundary. These dimensions distinguish changes that a component list obscures.

\begin{figure}[t]
\centering
\begin{tikzpicture}[x=0.84in,y=0.85in,
  cell/.style={rounded corners=2pt, draw=black!28, fill=white, line width=0.45pt,
    align=center, text width=0.96in, minimum height=0.29in, inner xsep=3pt,
    inner ysep=2.5pt, font=\scriptsize},
  lane/.style={align=left, text width=0.72in, font=\scriptsize},
  head/.style={align=center, text width=1.05in, font=\scriptsize\bfseries},
  blue/.style={cell, draw=cvblue!75, fill=lightblue},
  green/.style={cell, draw=cvgreen!75, fill=lightgreen},
  amber/.style={cell, draw=cvamber!75, fill=lightamber},
  red/.style={cell, draw=cvred!75, fill=lightred}
]
  \coordinate (xone) at (-1.80,0);
  \coordinate (xtwo) at (-0.35,0);
  \coordinate (xthree) at (1.10,0);
  \coordinate (xfour) at (2.55,0);
  \node[head] at (-1.80,0.90) {CosyVoice};
  \node[head] at (-0.35,0.90) {CosyVoice 2};
  \node[head] at (1.10,0.90) {CosyVoice 3};
  \node[head] at (2.55,0.90) {Qwen-Audio-\\3.0-TTS};
  \draw[cvblue, line width=1.15pt] (-2.27,0.68) -- (-1.33,0.68);
  \draw[cvgreen, line width=1.15pt] (-0.82,0.68) -- (0.12,0.68);
  \draw[cvamber, line width=1.15pt] (0.63,0.68) -- (1.57,0.68);
  \draw[cvred, line width=1.15pt] (2.08,0.68) -- (3.02,0.68);

  \node[lane] at (-2.72,0.42) {\textbf{R}\; what\\crosses};
  \node[lane] at (-2.72,0.02) {\textbf{O}\; who\\owns it};
  \node[lane] at (-2.72,-0.38) {\textbf{A}\; when it\\arrives};
  \node[lane] at (-2.72,-0.78) {\textbf{G}\; what can\\shape it};
  \foreach \y in {0.42,0.02,-0.38,-0.78}
    \draw[-{Latex[length=1.6mm]}, black!22, line width=0.55pt]
      (-2.25,\y) -- (3.18,\y);

  \node[blue]  at (-1.80,0.42) {supervised\\semantics};
  \node[cell]  at (-0.35,0.42) {full-use\\FSQ};
  \node[amber] at (1.10,0.42) {multitask +\\rewardable};
  \node[red]   at (2.55,0.42) {low-rate FSQ\\+ hidden state};

  \node[cell]  at (-1.80,0.02) {speaker in\\LM + FM};
  \node[green] at (-0.35,0.02) {speaker cues\\$\rightarrow$ FM};
  \node[cell]  at (1.10,0.02) {specialized\\roles};
  \node[red]   at (2.55,0.02) {stage-specific\\ownership};

  \node[cell]  at (-1.80,-0.38) {offline\\boundary};
  \node[green] at (-0.35,-0.38) {causal chunks\\+ bi-stream};
  \node[cell]  at (1.10,-0.38) {causal path\\retained};
  \node[red]   at (2.55,-0.38) {12.5-Hz\\planning};

  \node[cell]  at (-1.80,-0.78) {separate\\objectives};
  \node[cell]  at (-0.35,-0.78) {pretrain, then\\compose};
  \node[amber] at (1.10,-0.78) {DiffRO\\$\rightarrow$ LM};
  \node[red]   at (2.55,-0.78) {flow $\rightarrow$ LM\\+ staged RL};
\end{tikzpicture}
\caption{The interface-contract map. Rows unpack $\mathcal{I}=(R,O,A,G)$ and columns
follow the lineage. Color marks an explicit redesign; neutral cells show carried
context. The map is cumulative rather than a scalar score.}
\label{fig:contract-map}
\end{figure}
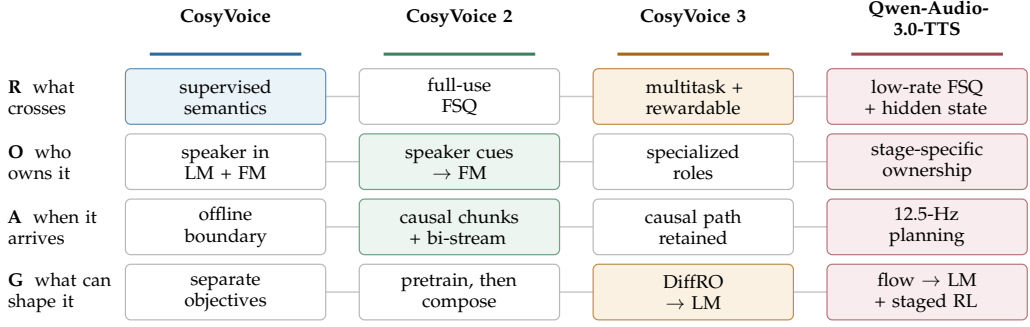

\paragraph{Operational definition of a bottleneck.}
We use \emph{bottleneck} as an analytical label when a release explicitly targets a
constraint and reports either a controlled ablation or a capability evaluation tied to
that intervention. Where no matched ablation exists, we describe relocation as an
interpretation rather than a measured causal effect. The definition is deliberately
local: a tokenizer ablation may identify a constraint for one configuration without
showing that it dominates at every data scale or deployment condition.

\paragraph{Rate is not capacity.}
For a tokenizer with frame rate $r$ and codebook size $K$, the nominal index budget is
\begin{equation}
 B_{\max}=r\log_2 K\quad\text{bits/s}.
 \label{eq:budget}
\end{equation}
This is an upper bound under uniform independent codes, not a measured entropy rate.
It is nevertheless useful for separating temporal rate from per-token capacity. A lower
frame rate shortens autoregressive sequences; increasing $K$ can partly recover the
representational loss without restoring the original sequence length.

\section{The speech-generation design space}
\label{sec:designspace}

The lineage sits between two historically separate views of speech synthesis. Codec
language models treat speech as discrete sequences and reuse autoregressive language
modeling machinery, as exemplified by VALL-E~\citep{wang2023vall-e}. Fully
non-autoregressive systems such as Voicebox~\citep{le2023voicebox},
E2-TTS~\citep{eskimez2024e2tts}, and F5-TTS~\citep{chen2024f5tts} generate continuous
acoustics through flow or diffusion-like objectives. MaskGCT~\citep{wang2024maskgct}
is also non-autoregressive, but operates in discrete token space through iterative
mask-and-predict refinement. Continuous autoregressive designs such as
DiTAR~\citep{jia2025ditar} instead model continuous patches. Seed-TTS explicitly
reports both autoregressive and non-autoregressive variants
\citep{anastassiou2024seedtts}.

These approaches are best organized by the object that carries long-range planning and
the mechanism used for local acoustic realization, rather than by a false binary between
``discrete'' and ``continuous.'' A hybrid assigns different time scales to different
modules: the \LM makes text-conditioned decisions with long-range consequences, while
the \FM reconstructs dense acoustics. Its planner must therefore emit units that are
predictable, sufficiently informative, temporally available, and useful to the renderer.
The decisive question is what information and optimization signals cross their boundary;
the rest of the paper follows how each term of that contract became limiting.

\section{Four bottleneck relocations}
\label{sec:acts}

\subsection{CosyVoice: from acoustic compression to supervised semantics}

The first CosyVoice identifies representation as the central problem. Conventional
neural codecs are optimized for waveform reconstruction, so their codes carry fine
acoustic variation that may be weakly determined by input text. Asking an autoregressive
\LM to predict such codes couples content planning to nuisance variation and lengthens
the search for a text-aligned latent space. CosyVoice instead inserts vector
quantization into a multilingual ASR encoder and trains a 4,096-entry codebook under
supervised recognition objectives~\citep{du2024cosyvoice}. The resulting supervised
semantic tokens are designed to preserve semantic information and improve alignment to
text.

The system then factorizes generation. The \LM predicts semantic tokens from normalized
text, prompt context, and an utterance-level speaker embedding. An optimal-transport
conditional flow-matching model maps those tokens to mel-spectrograms, additionally
conditioned on prompt acoustics and the speaker embedding, and a neural vocoder
reconstructs the waveform. Flow matching supplies a direct continuous path for details
that need not be serialized into the semantic token sequence
\citep{lipman2023flowmatching}. This architecture establishes the invariant that remains
throughout the lineage: \emph{plan compactly, render continuously}.

The gain is conceptual as well as empirical. Supervision aligns the token partition with
the text-to-speech task, shifting the first interface dimension $R$ from reconstruction
codes to content-oriented units. Yet the first design leaves three questions open:
whether the quantizer uses its finite vocabulary effectively, whether the computation is
causal enough for streaming, and whether speaker identity should be exposed to the
planner at all.

\subsection{CosyVoice 2: causal availability and factor ownership}

\cvtwo targets those questions with a coordinated set of changes
\citep{du2024cosyvoice2}. First, finite scalar quantization (FSQ)
\citep{mentzer2024fsq} replaces vector quantization. The paper reports a 6,561-code
space with full utilization, compared with 23\% utilization for a 4,096-entry VQ
baseline in the tokenizer evaluation. Second, the \LM is initialized from Qwen2.5-0.5B
and stripped of both a separate text encoder and its utterance-level speaker embedding.
Prompt text and speech tokens remain part of in-context learning, while explicit
speaker embedding and reference acoustics condition the \FM. Third, a bi-streaming
schedule interleaves text and speech tokens, while a chunk-aware causal \FM switches
between streaming and offline masks.

These interventions change $R$, $O$, and $A$ simultaneously. FSQ improves the usable
partition of the discrete interface. Removing the speaker embedding assigns linguistic
planning and explicit timbre conditioning more cleanly across the \LM and acoustic
model; it does not make the \LM independent of the prompt. Chunk-aware attention makes
local acoustic outputs available before the full utterance is known.

The modular ablation in \cvtwo is especially informative. On the reported Chinese,
English, and hard subsets, initializing from a pretrained \LM, removing its speaker
embedding, and replacing VQ with FSQ progressively reduce recognition error. The FSQ
step moves Chinese CER from 2.56 to 1.45, English WER from 3.81 to 2.57, and hard-set WER
from 9.66 to 6.83; speaker-similarity changes are comparatively small
\citep{du2024cosyvoice2}. The result supports two local conclusions: codebook
utilization was associated with content accuracy in this configuration, and removing
the utterance-level vector from the planner could improve content prediction without
removing explicit speaker conditioning from the system. The condition was relocated,
not eliminated.

The streaming ablation also separates module effects. Offline and streaming \LM/\FM
combinations preserve similar accuracy and similarity on the standard sets, while the
hard subset is more sensitive to streaming scheduling. This makes availability a
first-class interface property rather than a serving-layer afterthought: a causal
renderer is useful only if the planner's outputs arrive at compatible boundaries.

\subsection{CosyVoice 3: scale, coverage, and rewardable representations}

In the chronology of the series, \cvthree next targets data coverage and in-the-wild
generalization. Earlier published CosyVoice mixtures contain roughly 170,000 hours,
while CosyVoice~3 reports one million hours of generation training
data. It extends coverage across nine languages and broad Chinese
dialect variation, and scales the \LM from 0.5B to 1.5B parameters
\citep{du2024cosyvoice,du2024cosyvoice2,du2025cosyvoice3}. Crucially, scaling is
paired with a stronger tokenizer rather than treated as a substitute for representation
learning. A multitask voice encoder learns 25-Hz FSQ tokens from 530,000 hours spanning
ASR, language identification, speech emotion recognition, audio-event detection, and
speaker analysis. The same release scales a diffusion-transformer \FM from 100M to
300M parameters and makes HiFT causal, avoiding boundary blur from streaming
fade-in/fade-out \citep{funaudiollm2025cosyvoice3code}.

The tokenizer downstream study reports large gains from 3,000 to 170,000 hours, then
relates them to one-million-hour full-system results whose improvement begins to
plateau. Because the last point is not a matched tokenizer-only intervention, this is
descriptive evidence, not a controlled three-point curve. We interpret the trend as a
shift toward residual cases not fully addressed by likelihood training; it motivates,
but does not prove, the need for post-training.

\cvthree introduces differentiable reward optimization (DiffRO) for this role. Rather
than sampling full waveforms for every reward computation, DiffRO applies
Gumbel--Softmax relaxation to the token distribution and uses token-domain reward
models, allowing corrective gradients to reach the \LM directly
\citep{du2025cosyvoice3}. The method expands from recognition-oriented rewards to
multitask rewards, including expressive attributes. In interface terms, the discrete
plan is not merely compressed and predicted; it is made \emph{rewardable}. The change
does not yet give acoustic reconstruction losses access to the planner, but it broadens
$G$ from token likelihood alone to task-specific token-domain feedback.

\subsection{Qwen-Audio-3.0-TTS: capacity and cross-module coordination}

\system inherits the same high-level decomposition but confronts two pressures
\citep{xiang2026qwenaudio}. Production latency favors fewer autoregressive steps, so its
supervised tokenizer reduces the frame rate from 25 to 12.5~Hz. At the same time, broad
control, long-form generation, speaker fidelity, and robustness demand information that
may not survive a narrower token-only channel. The solution is not to reverse course to
high-rate acoustic tokens. It combines a higher-capacity FSQ space, containing
$3^{10}=59{,}049$ possible indices, with continuous \LM hidden states as the \FM
conditioning interface in place of discrete token embeddings.

Table~\ref{tab:tokenizer} reproduces the within-paper tokenizer comparison reported by
\citet{xiang2026qwenaudio}. Halving the rate while keeping $K=6{,}561$ degrades both
content consistency and similarity. Enlarging $K$ to 19,683 or 59,049 recovers the loss,
with the largest codebook providing the lowest reported content error among the
12.5-Hz variants. Equation~\ref{eq:budget} clarifies why the result is not a contradiction:
25-Hz, 6,561-way codes have a nominal index budget of about 317 bits/s, whereas 12.5-Hz,
59,049-way codes use about 198 bits/s. The latter has fewer autoregressive steps and a
larger choice per step; supervision and code utilization determine how effectively that
budget is used.

\begin{table}[t]
\centering
\small
\setlength{\tabcolsep}{3.7pt}
\begin{tabular}{lrr cc cc cc}
\toprule
& \multicolumn{2}{c}{\textbf{Interface}} &
\multicolumn{2}{c}{\textbf{test-zh}} &
\multicolumn{2}{c}{\textbf{test-en}} &
\multicolumn{2}{c}{\textbf{test-hard}} \\
\cmidrule(lr){2-3}\cmidrule(lr){4-5}\cmidrule(lr){6-7}\cmidrule(lr){8-9}
\textbf{Tokenizer} & \textbf{$K$} & \textbf{Hz} &
\textbf{CER$\downarrow$} & \textbf{SIM$\uparrow$} &
\textbf{WER$\downarrow$} & \textbf{SIM$\uparrow$} &
\textbf{CER$\downarrow$} & \textbf{SIM$\uparrow$} \\
\midrule
CosyVoice 3 & 6,561 & 25.0 & 1.45 & 80.60 & 2.57 & 73.60 & 6.83 & 77.60 \\
\system & 6,561 & 12.5 & 2.59 & 72.44 & 3.21 & 61.64 & 7.94 & 69.78 \\
\system & 19,683 & 12.5 & 1.48 & \textbf{83.25} & 2.56 & \textbf{77.58} &
6.70 & \textbf{80.85} \\
\system & 59,049 & 12.5 & \textbf{1.23} & 83.09 & \textbf{2.37} & 77.49 &
\textbf{6.68} & 80.61 \\
\bottomrule
\end{tabular}
\caption{Within-paper tokenizer comparison from \system on SEED-TTS-Eval
\citep{xiang2026qwenaudio}. CER/WER and SIM are percentages. Bold marks the best
12.5-Hz result in each evaluation column; the CosyVoice~3 row is the 25-Hz reference
reported in the same ablation table. The source labels \emph{test-hard} as CER; the
CosyVoice~2 modular table labels the same 6.83 baseline as WER.}
\label{tab:tokenizer}
\end{table}

\begin{figure}[t]
\centering
\begin{tikzpicture}[
  font=\small,
  mod/.style={draw=black!65, rounded corners=2pt, align=center, minimum height=0.42in,
    text width=0.82in, fill=white},
  data/.style={draw=black!55, rounded corners=8pt, align=center, minimum height=0.30in,
    text width=0.90in, fill=lightgray},
  arr/.style={-{Latex[length=2mm]}, line width=0.75pt, draw=black!70},
  grad/.style={-{Latex[length=2mm]}, line width=0.85pt, dashed, draw=cvred}
]
  \node[font=\small\bfseries, anchor=west] at (-2.55in,0.92in)
    {CosyVoice 1--3: token-conditioned LM--FM boundary};
  \node[mod] (lma) at (-2.02in,0.43in) {\LM\\planner};
  \node[data] (za) at (0,0.43in) {token IDs $z$};
  \node[mod] (fma) at (2.02in,0.43in) {\FM\\renderer};
  \draw[arr] (lma) -- (za);
  \draw[arr] (za) -- (fma);
  \node[font=\small, text=black!65] at (0,-0.02in)
    {the discrete plan is the LM--FM bridge};

  \node[font=\small\bfseries, anchor=west] at (-2.55in,-0.54in)
    {\system: token target plus hidden-state conditioning};
  \node[mod] (lmb) at (-2.02in,-1.15in) {\LM\\planner};
  \node[data] (zb) at (0,-0.88in) {token IDs $z$};
  \node[data, fill=lightred, draw=cvred] (hb) at (0,-1.42in) {hidden states $h$};
  \node[mod] (fmb) at (2.02in,-1.15in) {\FM\\renderer};
  \draw[arr] (lmb.north east) -- (zb.west);
  \node[font=\small, text=black!65] (tokloss) at (1.02in,-0.88in)
    {token loss};
  \draw[arr] (zb.east) -- (tokloss.west);
  \draw[arr, draw=cvred] (lmb.south east) -- (hb.west);
  \draw[arr, draw=cvred] (hb.east) -- (fmb.south west);
  \draw[grad] ([yshift=-2pt]fmb.south west) to[bend left=21]
    node[below, font=\small, text=cvred] {flow-loss gradient reaches the planner}
    ([yshift=-2pt]lmb.south east);
\end{tikzpicture}
\caption{The interface transition, with prompt and speaker side conditions omitted.
Earlier systems condition the \FM on embeddings of the discrete plan. In \system,
semantic tokens remain a prediction target, while continuous \LM hidden states replace
token embeddings as \FM conditioning and carry the upstream acoustic gradient during
joint training.}
\label{fig:interface}
\end{figure}
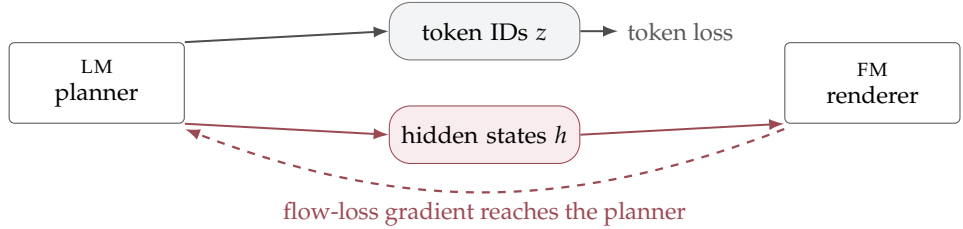

Codebook scaling improves the discrete prediction target. During joint training,
\system conditions the \FM on continuous \LM hidden states instead of discrete token
embeddings, while retaining semantic token prediction as a training objective. The
flow loss can therefore update upstream representations through $h_{1:T}$.
This conditioning mechanism is not claimed as a first: the \system report explicitly
aligns it with JoyVoice~\citep{yu2025joyvoice}. The lineage-level contribution analyzed
here is its insertion into a progressive schedule initialized from an independently
pretrained CosyVoice-style cascade.
Figure~\ref{fig:interface} contrasts this two-branch training graph with the earlier
token-conditioned cascade. Discrete tokens remain an interpretable semantic scaffold
and autoregressive sampling target; hidden states carry the renderer-conditioning
signal and expose the planner to end-to-end acoustic pressure.

This is a change to $R$ and $G$, not a rejection of modularity. Independent pretraining
still supplies stable components, and later stages selectively freeze one module while
specializing the other. The architecture is best understood as \emph{softening the
boundary}: the system keeps explicit roles but permits richer information and carefully
scheduled gradients to cross between them.

\section{From scaling to staged alignment}
\label{sec:stages}

A richer interface creates an optimization problem. Training all components
end-to-end from the start would force the \LM to learn semantic planning while its
conditioning target and downstream gradients are moving. Keeping the cascade frozen
forever would preserve stability but leave interface mismatch unresolved. \system
therefore uses five progressive stages, shown in Table~\ref{tab:stages}.

\begin{table}[t]
\centering
\small
\setlength{\tabcolsep}{4pt}
\begin{tabularx}{\textwidth}{P{0.30in} P{1.11in} P{1.25in} Y Y}
\toprule
& \textbf{Stage} & \textbf{Trainable path} & \textbf{Primary signal} &
\textbf{Bottleneck targeted} \\
\midrule
1 & Independent pretraining & \LM and \FM separately &
token likelihood; flow matching &
stable semantic planning and acoustic reconstruction \\
2 & Joint training and quality annealing & hidden-state-conditioned \LM--\FM &
token plus flow losses; curated late mixture &
interface mismatch, fidelity, and controllability \\
3 & \LM reinforcement learning & \LM; downstream synthesis fixed &
GRPO plus selective DiffRO &
content, duration, diversity, prosody, and dialect behavior \\
4 & Acoustic robustness & \FM; \LM frozen &
degraded-prompt augmentation &
voice fidelity under noise, reverb, devices, and packet loss \\
5 & \FM reinforcement learning & \FM; \LM fixed &
group-relative terminal acoustic rewards &
speaker similarity, intelligibility, and perceptual quality \\
\bottomrule
\end{tabularx}
\caption{The five-stage training curriculum in \system
\citep{xiang2026qwenaudio}. Each stage begins from the preceding checkpoint and
activates the module closest to the current residual error.}
\label{tab:stages}
\end{table}

\paragraph{Stage 1: establish the factorization.}
The \LM learns text-to-token planning while the \FM learns token-to-acoustic
reconstruction. This recovers the modular initialization validated in earlier releases
and prevents low-level acoustic losses from destabilizing an unformed semantic space.

\paragraph{Stage 2: couple after competence.}
The pretrained modules are connected through continuous hidden states and optimized
with token and flow objectives. Training begins with broad data, then anneals toward a
cleaner and more expressive subset. The order matters: broad coverage is learned before
the data distribution narrows toward quality. Joint optimization turns the interface
from a fixed handoff into a learned coordination layer.

\paragraph{Stage 3: optimize the planner in its native domain.}
With the renderer fixed, group relative policy optimization (GRPO) supplies
sequence-level rewards for content, duration, diversity, and prosody, while a filtered
DiffRO branch adds token-level corrective gradients. Computing these signals before
waveform synthesis reduces rollout cost. The stage illustrates a general principle:
when failures can be diagnosed in the plan, optimize the planner without repeatedly
invoking the expensive renderer.

\paragraph{Stage 4: isolate prompt robustness in the renderer.}
The \LM is frozen while the \FM is trained with noisy, reverberant, bandwidth-limited,
device-corrupted, far-field, blocked-microphone, packet-loss, and compound prompt
degradations. The assignment follows factor ownership. The semantic plan need not
imitate recording artifacts; the renderer must recover voice identity and clean
acoustics from them.

\paragraph{Stage 5: align acoustic outcomes.}
Finally, \FM reinforcement learning uses stochastic sampling for on-policy exploration
and group-relative terminal rewards combining speaker-verification similarity, ASR
intelligibility, and DNSMOS quality, with the \LM fixed. The separation is intended to
reduce credit-assignment
ambiguity:
plan-level post-training precedes renderer-level post-training, so the two stages target
different residuals.

The five stages are more than an implementation schedule. They are an optimization
counterpart to the interface contract. Ownership determines which component should
change; availability determines the serving constraint; representation determines what
can be expressed; and gradient reach determines when end-to-end feedback is safe. In
this view, ``pretraining, joint training, and reinforcement learning'' are not a bag of
techniques but a sequence that progressively expands the trainable system.

\section{Transferable design principles}
\label{sec:principles}

\paragraph{Diagnose the residual, not the release.}
The correct unit of analysis is a failure mode under a deployment condition. Content
errors with stable similarity implicate representation or planning. Similarity loss
under degraded prompts implicates acoustic conditioning. Latency with unchanged
quality implicates availability or token rate. Improvements should be evaluated against
the metric family closest to the intervention.

\paragraph{Treat tokens as a rate--capacity--supervision trade-off.}
Frame rate alone is not a measure of information content, and codebook size alone is not
effective capacity. Report at least $(r,K)$, code utilization or empirical entropy,
intrinsic content probes, and downstream synthesis metrics. The Qwen tokenizer ablation
shows why: lowering $r$ at fixed $K$ hurts, while increasing per-step capacity can
recover much of the loss. A useful tokenizer is one the planner can predict and the
renderer can exploit, not merely one that reconstructs audio.

\paragraph{Place explicit conditions at the module that consumes them.}
The two papers provide complementary evidence rather than a universal proof. \cvtwo
reports that removing the utterance-level speaker vector from the \LM reduces content
errors while the \FM retains speaker conditions. \system freezes the \LM and trains the
\FM for degraded-prompt robustness. Together they motivate the hypothesis that explicit
acoustic conditions should enter no earlier than their function requires, which may
improve credit assignment and reduce leakage between content and acoustics.

\paragraph{Preserve discrete scaffolds while relaxing hard bottlenecks.}
The two-branch training graph reconciles two needs. Discrete tokens provide compact
causal decisions and a convenient domain for ASR-like supervision and reward.
Continuous hidden states provide the conditioning signal for acoustic realization and
carry the flow-loss gradient upstream. Future systems need not use the same
representation for semantic prediction and acoustic conditioning.

\paragraph{Couple progressively and specialize selectively.}
Independent pretraining provides stable roles, joint training addresses interface
mismatch, and frozen-module stages target residuals with clearer ownership. This
pattern is likely to transfer to other modular generators in which a slow semantic
planner drives a dense continuous renderer.

\section{Evaluation protocol for future lineages}
\label{sec:protocol}

The source papers provide strong local evidence but not a fully matched longitudinal
experiment. A future lineage study should freeze evaluation artifacts and measure
interventions along four axes.

\textbf{Representation} should include token rate, vocabulary size, utilization,
empirical entropy, ASR probes, speaker leakage probes, and reconstruction quality.
\textbf{Ownership} should ablate prompt and speaker variables at the \LM and \FM
separately. \textbf{Availability} should report first-packet latency, real-time factor,
chunk size, look-ahead, and offline--streaming quality deltas on the same checkpoint.
\textbf{Gradient reach} should compare independent, joint, stop-gradient, and staged
training while controlling data and parameter count.

All generations should use fixed text, prompts, languages, random seeds, ASR systems,
speaker encoders, and perceptual raters. Results should be stratified by ordinary,
hard-content, cross-lingual, expressive-control, long-form, and degraded-prompt
conditions. Factorial experiments are especially valuable: tokenizer rate should be
crossed with vocabulary size; speaker conditioning should be crossed with module
location; and joint gradients should be crossed with initialization. Reporting only the
final system confounds bottleneck removal with data and scale.

Finally, failures should be routed before they are aggregated. A single mean score can
hide whether a model repeats content, drifts in speaker, violates an instruction, or
fails only under streaming. The interface contract supplies a diagnostic map from each
failure family to a plausible module and makes ablations falsifiable.

\section{Limitations, responsible use, and conclusion}

This paper is a retrospective synthesis of published systems, not a new matched
benchmark. Public papers document different checkpoints and evaluation stacks; our
bottleneck sequence is therefore a mechanistic interpretation constrained by local
ablations, not a claim that one variable explains every release-level gain. The
framework also emphasizes the \LM--\FM boundary and gives less attention to data
governance, front-end normalization, vocoder engineering, serving infrastructure, and
human preference measurement. These factors can independently become dominant in
production.

High-fidelity zero-shot synthesis also creates risks of impersonation, fraud,
unauthorized voice cloning, and scalable misinformation. Deployment requires
consent-aware enrollment, provenance or watermarking where effective, access controls,
abuse monitoring, multilingual red-teaming, and synthetic-media disclosure.
Degraded-prompt robustness deserves careful authorization because it lowers the quality
threshold for voice enrollment.

Across four generations, the stable idea is compact semantic planning followed by
continuous acoustic rendering; progress came from changing their contract. CosyVoice
aligned representation, CosyVoice~2 made it causal and clarified ownership,
CosyVoice~3 scaled and post-trained the plan, and \system expanded the interface and
coordinated optimization. Modular generators improve by identifying which boundary
property limits them---what crosses, who owns it, when it arrives, and how gradients
reach it.

\section*{LLM-use disclosure}

An LLM assisted organization, editing, and LaTeX/diagram code. The authors chose the
thesis, verified cited claims and numbers, and remain responsible for the paper.

\clearpage
\bibliography{references}
\bibliographystyle{colm2026_conference}

\end{document}